# The Narayana Universal Code

Krishnamurthy Kirthi and Subhash Kak
Oklahoma State University, Stillwater

**Abstract**
This paper presents a method of universal coding based on the Narayana series. The rules necessary to make such coding possible have been found and the length of the resulting code has been determined to follow the Narayana count.

**Keywords:** Fibonacci coding, Elias coding, GH sequences, Narayana sequences, data compression.

1. Introduction

A universal code maps positive integers which represent the source messages into codewords of different lengths. The codeword elements are a set of digits that are constructed according to a specified rule, and they may be binary. There are various universal codes including the Elias codes, the Fibonacci universal code, Levenshtein coding and non-universal codes including unary coding, Rice coding, Huffman coding and Golomb coding [1]-[4]. If one were to represent numbers as sum of two prime numbers using Goldbach conjecture, inverse sequence may also sequences may also be used to construct a universal code [5].

The simplest of Elias codes is the gamma code in which the binary representation of the source code is preceded by [$\log_2 x$] zeroes indicate codeword for any natural number x, where $x \in N = \{1, 2, 3,..\}$. The time requirement for compression and decompression algorithms for cases where decompression time is a critical issue, is advantageous in this coding [6],[7].

The Fibonacci code has a useful property of easy recovery of data from damaged bit stream in comparison with other universal codes. The performance of Fibonacci universal code is better than that of Elias coding [8]. Fibonacci and GH universal codes are obtained based on Zeckendorf representation. In this representation, every positive integer can be represented uniquely as a sum of non-adjacent Fibonacci numbers [9],[10]. This helps in unique representation of codewords without two consecutive 1s, and this may be used for generalization of any coding.

Narayana (short for Narayana Pandit) wrote his famous book Gaṇita Kaumudi in 1356. His eponymous sequences [11],[12],[13], which are related to Fibonacci and GH sequences, have potential applications in cryptography and data coding. The properties of Fibonacci sequences together with applications in cryptography and coding have been presented in several studies [14]-[18]. Here, we present a variant of Fibonacci universal code based on Narayana series with the help of a representation procedure that leads to the Narayana universal code.

2. **Narayana sequence**

The Narayana sequence is derived from the following problem that was proposed by Narayana: "A cow gives birth to a calf every year. In turn, the calf gives birth to another calf when it is three years old. What is the number of progeny produced during twenty years by one cow?" We assume that we begin with a new-born calf, who is shown in the first row of the matrix below. After three years, in each successive year, there is a new calf born to this one and additional calves are born to the 3-year or older calves, leading o second and additional rows in the matrix every 3 steps. This may be represented in the matrix below:

Table 1. Generation of the Narayana sequence

| | | | | | | | | | | | | |
|---|---|---|---|---|---|---|---|---|---|---|---|---|
| 1 | 1 | 1 | 1 | 1 | 1 | 1 | 1 | 1 | 1 | 1 | 1 | 1 … |
| | | | 1 | 2 | 3 | 4 | 5 | 6 | 7 | 8 | 9 | 10… |
| | | | | | 1 | 3 | 6 | 10 | 15 | 21 | 28… |
| | | | | | | | | 1 | 4 | 10 | 20… |
| | | | | | | | | | | | 1… |
| | | | | | | | | | | | | …. |
| 1 | 1 | 1 | 2 | 3 | 4 | 6 | 9 | 13 | 19 | 28 | 41 | 60… |

The last row that adds up the numbers in the previous rows represents the count of the Narayana sequence. This sequence is the sum of previous term and term 2 places before. It is is given by

$$1,1,1,2,3,4,6,9,13,19,28, \ldots \quad . \tag{1}$$

The (k+1)st term of the Narayana series may be defined as:

$$N(k+1) = N(k) + N(k-2) \tag{2}$$

with $N(0) = N(1) = N(2) = 1$ and $k \geq 2$. A more general Narayana sequence $N_a(n)$ is given by

$$a, b, c, a+c, a+b+c, a+b+2c, 2a+b+3c, 3a+2b+4c, \text{ and so on, with } a=1, b=2 \text{ and } c=3 \tag{3}$$

Consider the ratio of two consecutive terms in Narayana series. In the limit where n goes to infinity, we have

$$\lim_{n\to\infty}\left(\frac{N_a(n+1)}{N_a(n)}\right) = 1 + \lim_{n\to\infty}\left(\frac{N_a(n-2)}{N_a(n)}\right) \quad (4)$$

Equation (4) may be written as,

$$\lim_{n\to\infty}\left(\frac{N_a(n+1)}{N_a(n)}\right) = 1 + \lim_{n\to\infty}\left(\frac{N_a(n-2)}{N_a(n-1)}\right) * \lim_{n\to\infty}\left(\frac{N_a(n-1)}{N_a(n)}\right) \quad (5)$$

With $\lim_{n\to\infty}\left(\frac{N_a(n+1)}{N_a(n)}\right) = L$, we obtain the equation $L^3 - L^2 - 1 = 0$. This leads to the following theorem:

**Theorem 1.**
The real positive solution of equation $L^3 - L^2 - 1 = 0$ characterizes the relation between two consecutive terms in Narayana sequence, and the Narayana ratio approaches 1.4655712318767669…

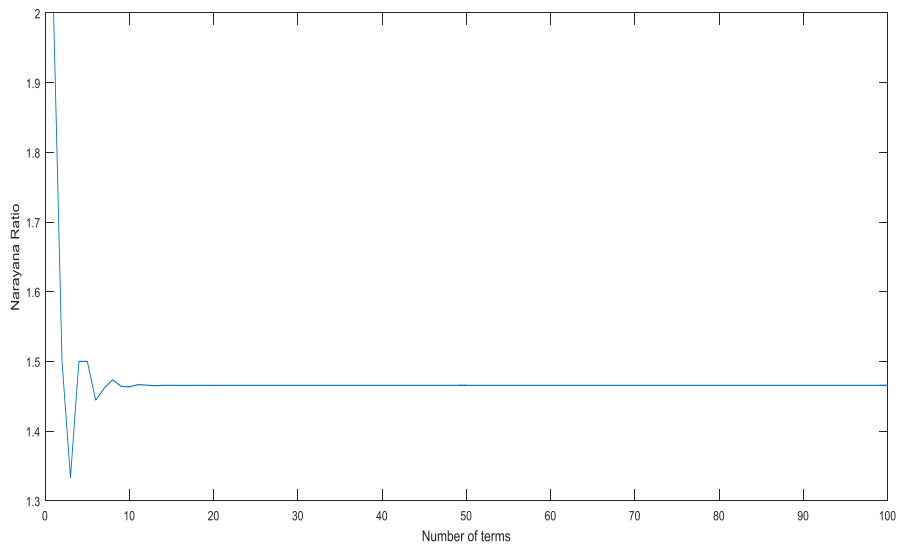

Figure 1. Ratio between first 100 consecutive terms of the Narayana series

This constant of 1.4655712318767669 may be termed the Narayana ratio.

### 3. The Narayana universal code

To generate Narayana code as a generalization of Fibonacci universal code, we need to be able to map any given positive integer representing source code into variable length codeword in a manner used earlier by Thomas [1].

Consider Narayana series N(k) given by 1,1,1,2,3,4,6,9,… for generating variable length source code for any given positive integer. Since the series contains 3 consecutive 1s, we obtain various codewords for any given positive integer and the resulting codeword fails to comply with the requirement of universal coding.

Hence, to it is essential to modify Narayana series N(k) for k terms such and we define a new series J(k) as N(k+2)=J(k). The series N(k) is mapped to the J series as shown in the Table 2.

Table 2. Mapping of N series into J series

| Narayana Series N(k) | | J(k) Series |
|---|---|---|
| 1 | N(0) | |
| 1 | N(1) | |
| 1 | N(2) | J(0) |
| 2 | N(3) | J(1) |
| 3 | N(4) | J(2) |
| 4 | N(5) | J(3) |
| 6 | N(6) | J(4) |
| 9 | N(7) | J(5) |
| 13 | N(8) | J(6) |
| 19 | N(9) | J(7) |
| 28 | N(10) | J(8) |

The conditions for unique representation of Narayana universal code in terms of J series to obtain binary set of codewords are now presented.

**Rule 1:** For a given positive integer *n*, construct a vector $A(n)$ such that $A(n)_i = J(i)$, $i=0,1,\ldots,d$, where $J(d)$ is the largest number of J series less than or equal to *n*. A vector $B(n)$ of binary digits with dimension d is constructed such that

$$A(n)^T B(n) = n \text{ and } B(n)_d = 1. \qquad (6)$$

The codeword $NB(n)$ for the positive integer *n* is defined by a vector with dimension d+1, where $NB(n)_k = B(n)_k$ for $1 \leq k \leq d$, and $NB(n)_{d+1} = 1$.

Consider an example of constructing codeword for integer 10. Since $J(5) = 9$ is the largest number of J series less than or equal to 10, vectors $A(n), B(n)$ and $NB(n)$ are given as (*d*=5 in this example):

$$A(n) = \begin{pmatrix} 1 \\ 2 \\ 3 \\ 4 \\ 6 \\ 9 \end{pmatrix}, B(n) = \begin{pmatrix} 1 \\ 0 \\ 0 \\ 0 \\ 0 \\ 1 \end{pmatrix}, NB(n) = \begin{pmatrix} 1 \\ 0 \\ 0 \\ 0 \\ 0 \\ 1 \\ 1 \end{pmatrix} \qquad (7)$$

While the recursive nature of Narayana series allows to have more than one representation for some integers using above scheme, $B(n)$ is chosen not to have two consecutive ones. In the above example, integer 10 can be represented as $J(3) + J(4)$ by $B(10) = (0\ \ 0\ \ 0\ \ 1\ \ 1)^T$ or using Zeckendorf representation as $J(0) + J(5)$ by $B(10) = (1\ \ 0\ \ 0\ \ 0\ \ 0\ \ 1)^T$. Since $NB(10)_{d+1} = NB(10)_d = 1$, consecutive ones occur only at the termination of codeword $NB(n)$, when Zeckendorf representation is chosen. Thus, the prefix conditions for unique representation of codeword are found to be:

**Rule 2:** If the source code to be represented is a term in Narayana series, the codeword consists of binary set with all zeroes followed by two consecutive ones at the termination.

**Rule 3:** If the source code to be represented is not a term in Narayana series, the codeword consists of binary representation of summation of two or more terms in Narayana sequence such that $A(n)^T B(n) = n$ which includes two consecutive ones at the termination as a part of Zeckendorf representation. We consider codeword which can be represented by summation of least number of terms in Narayana series.

Consider a general Narayana sequence which is given by *{a, b, c, a+c, a+b+c, a+b+2c, 2a+b+3c, and so on}*. Let *a+c=d, a+b+c=e, a+b+2c=f, 2a+b+3c=g* in the above sequence which represents *{a, b, c, d, e, f, g, and so on}*. Here, *g* is obtained by summation of *f* and *d* (that is, *g=d+f*) which in turn are summations of *e* and *c* and *c* and *a* respectively. Since any term of Narayana series is obtained by summation of two different terms in the sequence and codeword for any given positive integer is binary representation of sum of two or more terms in the sequence, the codeword obtained will be unique in Zeckendorf representation. Therefore, we are led to the following result:

**Theorem 2.**
The variable length codeword obtained in Zeckendorf representation for any given positive integer *n*, which represents the source message, is unique.

Table 3 provides codewords for first 15 natural numbers which contain source messages.

Table 3. Narayana code mapping for the numbers 1 through 15

| $n$ | Representation in terms of J series | Binary Representation in terms of J series | Narayana Code | Number of bits required for representation of Narayana Code |
|---|---|---|---|---|
| 1 | $J(0)$ | 1 | 11 | 2 |
| 2 | $J(1)$ | 01 | 011 | 3 |
| 3 | $J(2)$ | 001 | 0011 | 4 |
| 4 | $J(3)$ | 0001 | 00011 | 5 |
| 5 | $J(0)+J(3)$ | 1001 | 10011 | 5 |
| 6 | $J(4)$ | 00001 | 000011 | 6 |
| 7 | $J(0)+J(4)$ | 10001 | 100011 | 6 |
| 8 | $J(1)+J(4)$ | 01001 | 010011 | 6 |
| 9 | $J(5)$ | 000001 | 0000011 | 7 |
| 10 | $J(0)+J(5)$ | 100001 | 1000011 | 7 |
| 11 | $J(1)+J(5)$ | 010001 | 0100011 | 7 |
| 12 | $J(2)+J(5)$ | 001001 | 0010011 | 7 |
| 13 | $J(6)$ | 0000001 | 00000011 | 8 |
| 14 | $J(0)+J(6)$ | 1000001 | 10000011 | 8 |
| 15 | $J(1)+J(6)$ | 0100001 | 01000011 | 8 |

In order to decode the codeword, remove the last 1 in the codeword and assign the remaining bits with the values 1,2,3,4,6,9,13,19,… which are terms of the Narayana series (Narayana number) and add. Thus, the Narayana code can be used to encode any positive integer, which could be a portion of signal with source messages contained in it.

*Note.* A variant of Narayana coding scheme can be obtained by defining second order variant Narayana sequence, $VN_a(n)$, such that $b = 3 - a$ and $c = 1 - a$. This yields $VN_a(0) = a$ ($a \in Z$), $VN_a(1) = 3 - a$, $VN_a(2) = 1 - a$ and for $n \geq 3$, $VN_a(n) = VN_a(n-1) + VN_a(n-3)$.

With the above definition, we obtain variant Narayana sequence $VN_{-2}(n)$, which starts with $a = -2$, as {-2,5,3,1,6,9,10,16,25,..}. However, certain codewords cannot be represented with the above definition since there is no Zeckendorf representation for integer 2 using the above sequence.

Similarly, we obtain $VN_{-1}(n)$ as {*-1,4,2,1,5,7,8,13,20,28,..*}, $VN_{-3}(n)$ as {*-3,5,4,1,6,10,11,17,27,38,..*} and there is no Zeckendorf representation for integers 3 and 15 using the sequence $VN_{-1}(n)$ and integers 2,13 and 19 cannot be represented using sequence $VN_{-3}(n)$. ∎

Although, codes obtained through variant Narayana sequence are not capable for encoding certain positive integers, they could be used for portions of source messages which they are able of encode. But, variant Narayana sequences cannot be considered for universal coding.

Figure 2 presents the number of bits required for representation of codewords, obtained through Narayana universal coding, for first 1000 natural numbers. increases exponentially with the increase in number which may contain source message. XXX

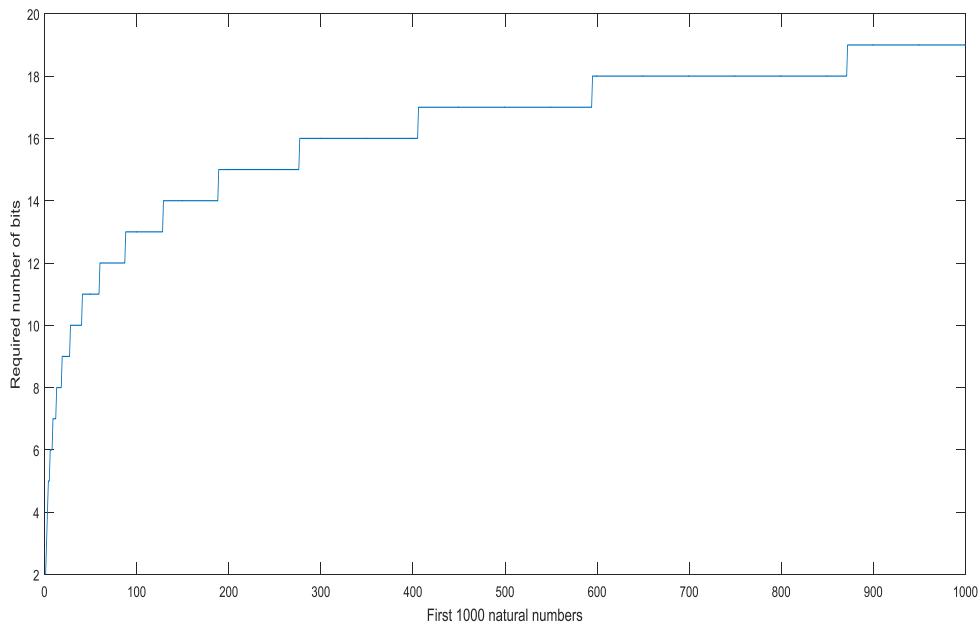

Figure 2. Required number of bits for codeword representation for first 1000 natural numbers

Let the number of elements with same number of bits, in the codeword representation for natural numbers obtained through Narayana universal coding, for repeated count greater than or equal to 2, be represented by the sequence b(x).

With the above definition, we have b(1)=2, b(2)=3, b(3)=4, b(4)=6 and so on. We see that sequence b(x) represents Narayana series.

**Theorem 3.**

The sequence *b(n)* which represents the number of elements with same number of bits for repeated count greater than or equal to 2, is also according to the Narayana series.

*Proof.* The above statement is true with respect to Rule 1 in which the codeword is defined by a vector with dimension *d+1*, where $NB(n)_k = B(n)_k$ for *1 ≤ k ≤ d*, and $NB(n)_{d+1} = 1$.

**Conclusion**

We have developed a method of using the Narayana series to generate a universal code. This method is based on the use of constraining rules that ensure that the coding is unique.